\documentclass[10pt,conference]{IEEEtran}

\usepackage{xcolor}

\usepackage{multirow} 
\usepackage{booktabs}
\usepackage{float}
\usepackage{tabularray}
\usepackage{siunitx}
\usepackage{lscape}
\usepackage{tikz}
\usepackage{tabularx}
\usepackage{cite}
\usepackage{amsmath,amssymb,amsfonts}
\usepackage{algorithmic}
\usepackage{graphicx}
\usepackage{textcomp}
\usepackage{xcolor}
\usepackage{booktabs}
\usepackage{subcaption}
\usepackage{float}
\usepackage{balance}
\IEEEoverridecommandlockouts
\begin{document}
%
%

\title{DECIFR: Domain-Aware Exfiltration of 
Circuit Information from Federated Gradient Reconstruction}

\author{\IEEEauthorblockN{Gijung Lee, Wavid Bowman, Olivia P. Dizon-Paradis, Reiner N. Dizon-Paradis, \\Ronald Wilson, Damon L. Woodard, Domenic Forte}
\IEEEauthorblockA{Florida Institute of National Security, University of Florida, Gainesville, Florida, USA\\
Email: \{lee.gijung, wavid.bowman, paradiso, reinerdizon, ronaldwilson\}@ufl.edu, \{dwoodard, dforte\}@ece.ufl.edu}


\vspace{-1em}
}

\maketitle


\begin{abstract}
Federated Learning (FL) is a promising approach for multiparty collaboration as a privacy-preserving technique in hardware assurance, but its security against adversaries with domain-specific knowledge is underexplored. This paper demonstrates a critical vulnerability where available standard cell library layouts (SCLL) can be exploited to compromise the privacy of sensitive integrated circuit (IC) training data. We introduce DECIFR, a novel two-stage Membership Inference Attack (MIA) that requires no auxiliary dataset. The attack employs a guided Gradient Inversion Attack (GIA) to reconstruct a client's training images from intercepted model updates. Our findings reveal that the fidelity of these reconstructions directly correlates with membership status, allowing an adversary to reliably distinguish members from non-members based on image quality. This work exposes a practical threat that overcomes the limitations of conventional attacks and underscores that standard FL protocols are insufficient for securing domains with extensive knowledge. We conclude that robust defenses are essential for the secure application of FL in hardware assurance.
\end{abstract}


\section{Introduction}
The globalized integrated circuit (IC) supply chain faces serious security threats such as hardware Trojans, IP piracy, IC counterfeiting, and IC overproduction \cite{rostami2014primer}, making hardware assurance a national priority as emphasized in the CHIPS Act~\cite{senate}. Deep learning models offer a powerful tool for automating security analysis, but they are hampered by a critical bottleneck: the slow and high cost acquisition of specialized SEM image data required for training. 

Federated Learning (FL) emerges as a promising solution, allowing organizations to collaboratively train models without sharing sensitive raw data. Despite its privacy-preserving design, FL remains vulnerable to attacks such as the membership inference attack (MIA), which determines if a specific data sample was in the training set~\cite{shokri2017membership, pyrgelis2017knock}. However, existing MIAs are often impractical in FL environments, as they require auxiliary datasets that a central server would not possess.

To address this gap, we introduce DECIFR, a novel MIA methodology targeting FL-trained Scanning Electron Microscopy (SEM) image segmentation models used for hardware assurance. Our key insight is that an adversary can leverage standard cell library layouts (SCLLs), which are accessible under NDA, to guide a gradient inversion attack (GIA). By intercepting a client's FL model update and using SCLLs as dummy labels in the GIA, DECIFR reconstructs the original training image. We hypothesize that an adversary can exploit this reconstruction process, as true training members will rebuild with markedly higher fidelity. This quality difference allows an attacker to reliably distinguish members from non-members. This work exposes a critical privacy vulnerability in the application of FL to hardware assurance and underscores the nuanced challenges of achieving robust hardware assurance.

Critically, membership inference in this domain poses a systemic threat by revealing not just data presence, but the underlying hardware characteristics (e.g., node or layer type) of the training set. This leakage of hardware-specific membership information acts as a critical enabler for physical attacks, as it provides additional information to accelerates reverse engineering~\cite{torrance2009state}, facilitates the identification of design weaknesses~\cite{courbon2020practical} to launch non-invasive attacks, and streamlines IP piracy~\cite{guin2014counterfeit}.
We make the following contributions:

\begin{itemize}
\item \textbf{Novel Data-Free Attack:} We propose the first data-free MIA for FL segmentation models, utilizing SCLLs to guide Gradient Inversion Attacks (GIA) without requiring private datasets.
\item \textbf{Hardware-Specific Inference:} We demonstrate that adversaries can infer specific hardware traits (e.g., layers, technology nodes) by analyzing differential reconstruction success using SCLL priors.
\item \textbf{Reconstruction-Based Detection:} We validate reconstruction fidelity as an effective metric for distinguishing members from non-members in hardware assurance.
\end{itemize}

\section{Background}
This section reviews the core concepts of FL and defines MIAs within this specific context.

\subsection{Federated Learning (FL)}
Federated Learning (FL) is a distributed framework that enables multiple entities to jointly train a global model while ensuring that their raw data remains local and private~\cite{Wen2023federated}. The FL protocol typically executes in iterative rounds: a central server distributes a global model, clients train it locally on private data, and then return only model updates to the server for aggregation~\cite{torkzadehmahani2022privacy}.

Common protocols include Federated Stochastic Gradient Descent (FedSGD) and Federated Averaging (FedAVG)~\cite{mcmahan2017communication}. While FedSGD transmits gradients directly, FedAVG requires clients to train for multiple epochs and share updated weights. Although FedAVG avoids direct gradient sharing, an adversary can still perform a Gradient Inversion Attack (GIA) by estimating gradients from the difference between the global model weights before and after local training. This estimation is the vulnerability exploited by our proposed attack.

\subsection{Membership Inference Attacks (MIAs)}
A Membership Inference Attack (MIA) is a privacy violation where an adversary determines if a specific data sample was part of a model's training set. In the context of hardware assurance, this vulnerability extends to the exposure of sensitive Intellectual Property (IP). For instance, consider a model trained on a dataset of Scanning Electron Microscopy (SEM) images. An adversary could probe this model by submitting distinct inputs, such as images of a 32nm metal layer versus a 32nm diffusion layer. If the model predicts the 32nm metal layer with disproportionately high confidence, the attacker can infer that this specific structure was likely included in the training set. Furthermore, by systematically testing various images from a known Standard Cell Library (SCL), an adversary can determine if the model was trained using images from that specific, proprietary Process Design Kit (PDK).

Pioneering research on MIAs proposed the shadow training technique \cite{shokri2017membership}, where an attacker trains surrogate models on ``shadow datasets" drawn from the same distribution as the victim's data. In the FL context, MIAs are broadly categorized into update-based and trend-based methods \cite{bai2024membership}.

\begin{itemize}
    \item \textbf{Update-based attacks} leverage the exchanged gradients or parameters. Specifically, these methods operate by analyzing the raw gradients and their temporal differences, or by applying shadow training techniques to approximate the target model's parameters in the federated environment. Early works like \cite{nasr2019comprehensive} require access to partial member data, while others exploit non-zero gradients in specific layers \cite{melis2019exploiting} or calculate gradient differences between rounds \cite{li2023effective}.

    \item \textbf{Trend-based attacks} infer membership by tracking metrics over the learning process. These techniques rely on analyzing the historical trajectory of indicators, such as prediction confidence or loss, to detect distributional differences between members and non-members. Recent approaches like CS-MIA \cite{gu2022cs} or data poisoning methods \cite{he2024enhance} effectively infer membership but often require the adversary to actively tamper with the training process or possess auxiliary datasets.
\end{itemize}

Crucially, most existing FL attacks assume the attacker has access to auxiliary data similar to the victim's or can actively poison the model. These assumptions are often impractical in hardware assurance, where high-quality domain data is scarce and proprietary. Our work addresses this gap by proposing a data-free attack that relies only on design knowledge (SCLLs).

\section{Methodology}
\label{sec:methods}
\begin{figure}[t]
\centering
\includegraphics[width=1\linewidth, height=0.5\linewidth, trim={0 1cm 0 0},clip]{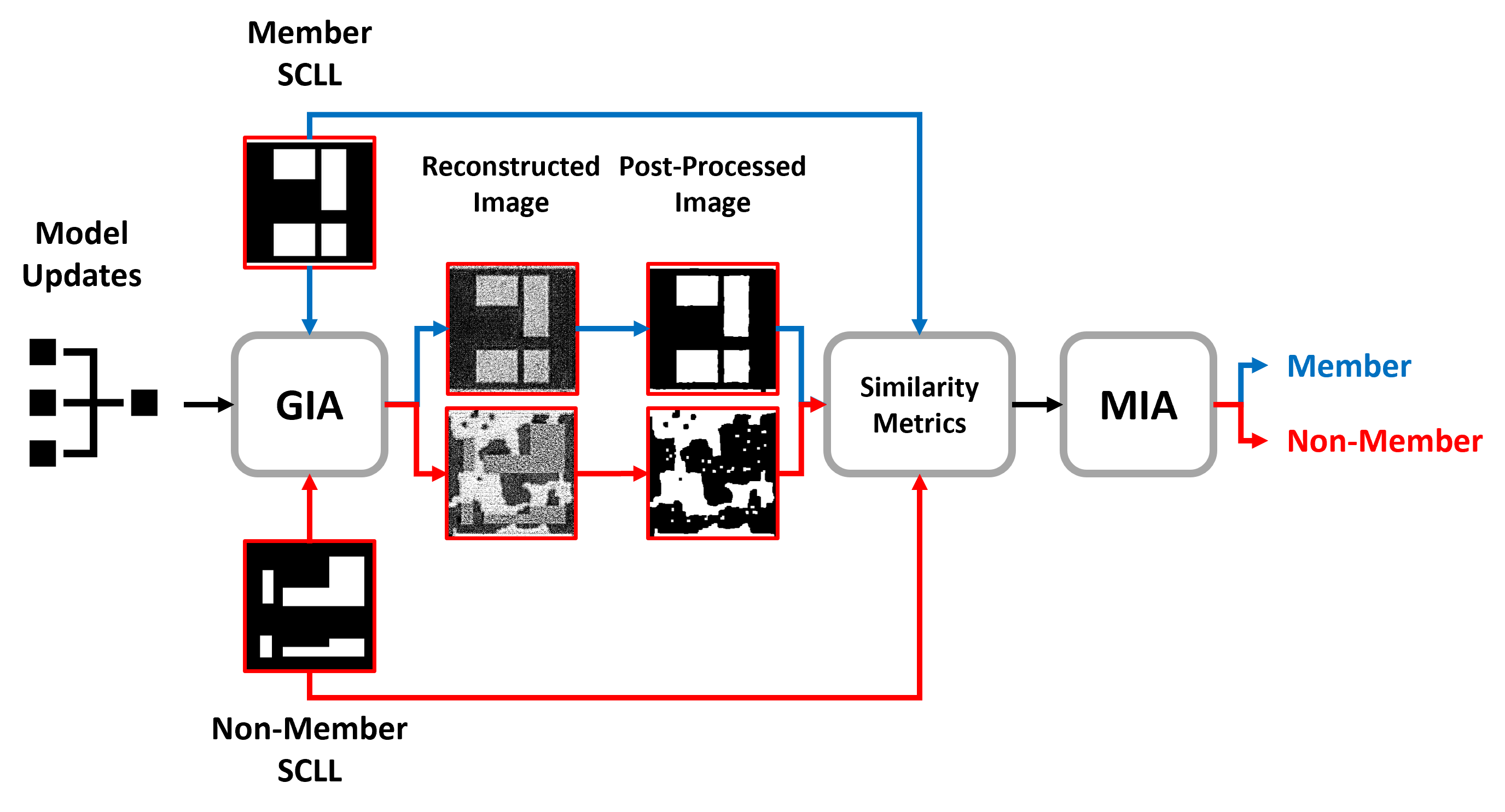}
\caption{The DECIFR Process: From intercepted model updates to membership inference via guided gradient inversion.}
\label{fig:MIA_process}
\end{figure}

In this study, we propose DECIFR, a novel MIA technique tailored for the FL environment. Unlike traditional attacks, DECIFR infers the presence of sensitive training data directly from model updates without relying on any auxiliary data from the target victim. Our experimental pipeline begins with a standard FL protocol where clients train a global segmentation model. As illustrated in Fig.~\ref{fig:MIA_process}, the attack itself launches from the intercepted model updates. In this first attack phase, an adversary performs a guided Gradient Inversion Attack (GIA) using Standard Cell Library Layouts (SCLLs) to reconstruct potential private training images from the updates. This is followed by the Membership Inference phase, where the fidelity of these reconstructions is analyzed to definitively determine if a specific hardware characteristic, such as a technology node or layer type, was present in the client's training data.

\subsection{Dataset and Environment Setup}

To evaluate our approach, we curated two distinct datasets of Standard Cell Library Layouts (SCLLs) derived from the 32nm and 90nm technology nodes within Synopsys’ Open Educational Design Kit (SAED). These datasets specifically isolate metal and diffusion layers. We applied a whitelist filtering process to extract a core set of logic gates (AND, NAND, OR, NOR, XOR, and XNOR) and essential supplementary cells (e.g., inverters, buffers), selecting only those that satisfied specific size constraints.

For the semantic segmentation task, we employed the REFICS\footnote{Link: https://trust-hub.org/\#/data/refics} tool \cite{wilson2021refics} to generate synthetic dataset of 141 SEM images and their corresponding ground-truth masks. The synthesis parameters were configured with a shot noise level of 20 and a dwelling time of 10 $\mu$sec/pixel to simulate realistic imaging conditions. Pixel intensity distributions were set with background and foreground means of 75 and 135, respectively, and a standard deviation of 20. All images were resized to a resolution of 256$\times$256 pixels. The segmentation model employs a 16-layer U-Net generator architecture ($\approx$54M parameters) adapted from Pix2Pix \cite{isola2017image}. The encoder utilizes 8 downsampling blocks with filter counts progressively increasing from 64 to 512, while the decoder utilizes 7 upsampling blocks with skip connections to recover spatial resolution.
All experiments were executed on a high-performance computing node equipped with an AMD EPYC ROME CPU (32GB RAM) and an NVIDIA B200 GPU (180GB vRAM).

\subsection{Threat Model}
We adopt an ``honest-but-curious'' threat model, where a participant (e.g., the central server) attempts to infer sensitive IP while strictly adhering to the FL protocol. This passive adversary monitors the training process but does not tamper with the model parameters or the aggregation process.
\begin{itemize}
    \item \textbf{Observable Information:} The adversary can access to model updates, such as weights or gradients, but lacks of local hyperparameters like the learning rate $\eta$.
    \item \textbf{No Auxiliary Data:} Unlike conventional MIAs, DECIFR operates without requiring a auxiliary dataset from the victim's distribution.
    \item \textbf{Domain Knowledge:} The adversary possesses a reference library of standard SCLLs, which can be obtained from PDKs.
\end{itemize}

\subsection{Phase 1: Federated Learning}
We simulated a FL environment for image segmentation consisting of a server and two clients, each possessing 50 image-mask pairs. A global U-Net \cite{ronneberger2015u} model was trained via the FedAvg algorithm over 200 communication rounds. In each round, clients executed 2 local epochs with a batch size of 10 and a learning rate of 0.01.

\subsection{Phase 2: Gradient Inversion}
\subsubsection{Gradient Extraction}
The attack begins at a target round (e.g., Round 200) by intercepting the client's update. Using the global weights ($W_{prev}$) and the received update ($W_{curr}$), the adversary derives the aggregated gradients ($\nabla \mathcal{L}$):
\begin{equation}
\nabla \mathcal{L} = \frac{W_{prev} - W_{curr}}{\eta}
\label{eq:grad_extraction}
\end{equation}
While we utilize a fixed learning rate ($\eta$) for experimental consistency, a sophisticated adversary could refine this parameter via grid search, selecting the $\eta$ that yields the most plausible reconstruction.

\subsubsection{Reconstruction via Guided GIA}
DECIFR boosts the GIA by using SCLLs as fixed dummy labels ($y'$). The adversary performs optimization twice: once guided by a metal SCLL ($y'_{metal}$) and once by a diffusion SCLL ($y'_{diff}$), iteratively updating a dummy image ($x'$) to minimize the composite loss function $L_{total}$:
\begin{equation}
    L_{total} = L_{grad} + \lambda_{tv} L_{tv} - \lambda_{dummy} L_{dummy}
    \label{eq:total_loss}
\end{equation}
The core Gradient Matching Loss ($L_{grad}$) enforces the dummy to match target gradients using a balanced mix ($\alpha=0.5$) of Mean Squared Error (MSE) and Cosine Similarity (CS):
\begin{align}
    L_{\text{grad}} ={}& \alpha \cdot \text{MSE}(\nabla \mathcal{L}_{dummy}, \nabla \mathcal{L}_{target}) \notag \\
    & + (1 - \alpha) \cdot \text{CS}(\nabla \mathcal{L}_{dummy}, \nabla \mathcal{L}_{target})
    \label{eq:grad_loss}
\end{align}
We apply Total Variation regularization ($L_{tv}$) with $\lambda_{tv}=0.001$. Uniquely, we introduce a negative Forward Pass Loss ($-L_{dummy}$), which penalizes the optimizer for merely fitting the fixed label. This constraint forces reliance on gradient matching, which is critical for recovering complex geometries like 32nm metal layers.

\subsection{Phase 3: Membership Inference}
\label{sec:MIA_processing}
The final phase determines membership based on reconstruction quality.
\begin{enumerate}
    \item \textbf{Post-Processing:} The raw reconstructions ($x'_{metal}$, $x'_{diff}$) are converted to binary masks using a pipeline of $5\times5$ Gaussian blur, Otsu's thresholding, and morphological opening/closing.
    \item \textbf{Scoring:} A similarity metric (Dice coefficient) is calculated between the binary mask and the corresponding SCLL.
    \item \textbf{Mean Threshold Attack:} We pool all similarity scores to calculate a global mean threshold $T$. If a reconstruction's score exceeds $T$, it is classified as a Member; otherwise, it is a Non-Member.
\end{enumerate}

\begin{figure*}[ht!] 
\vspace{-2mm}
\centering  
\includegraphics[width=1\linewidth, height=0.3\linewidth]{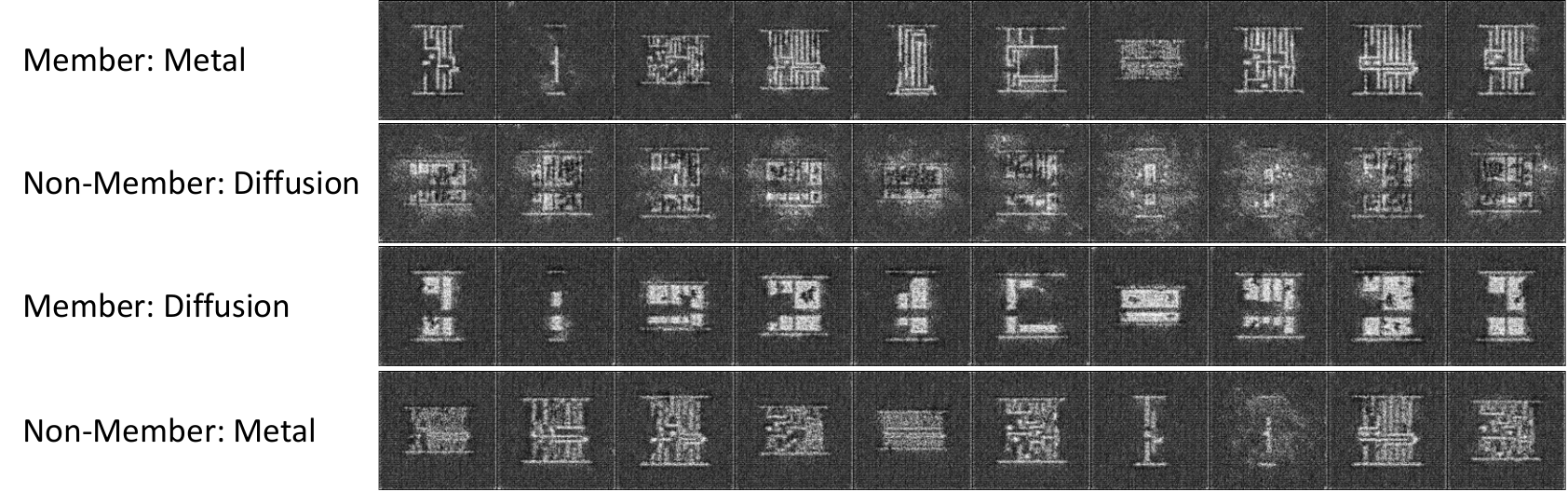} 
\caption{Representative GIA reconstructions illustrating the pronounced quality disparity in the metal-as-member case compared to the diminished gap in the diffusion-as-member case.}
\label{fig:reconstructions} 
\vspace{-3mm}
\end{figure*}

\section{Results}
\label{sec:results}
This section details the experimental evaluation of FL's vulnerability to MIAs via DECIFR. We structure our findings by first analyzing the primary inter-layer attack (metal vs. diffusion), followed by an ablation study of the negative Forward Pass Loss, and concluding with the challenging intra-layer scenario (32nm vs. 90nm).

\subsection{Analysis of Reconstructed Images}
The initial phase of DECIFR reconstructs the target's training data through guided GIA. Following our protocol, we generated two distinct reconstructions for every intercepted gradient: one anchored by a metal SCLL ($x'_{\text{metal}}$) and the other by a diffusion SCLL ($x'_{\text{diffusion}}$).

We observed a distinct asymmetry in reconstruction fidelity. For metal layer targets, the quality gap was pronounced: the member-guided output ($x'_{\text{metal}}$) achieved high fidelity, whereas the non-member version ($x'_{\text{diffusion}}$) was severely degraded. Conversely, for diffusion layer targets, this visual disparity narrowed, as both reconstructions retained relatively high quality. Fig.~\ref{fig:reconstructions} illustrates representative examples of these phenomena.

\begin{table}[t]
\centering
\caption{Results for Inter-Layer Dice Scores (at $\lambda_{\text{dummy}} = 0$)}
\label{tab:dice_scores}
\begin{tabular}{@{}llc@{}} \toprule Target Data & Guiding SCLL & Mean Dice Score \\ \midrule Metal & Metal & \textbf{0.7233} \\ Metal & Diffusion & 0.5876 \\ Diffusion & Diffusion & \textbf{0.8204} \\ Diffusion & Metal & 0.7130 \\ \bottomrule
\end{tabular}
\vspace{-3mm}
\end{table}

\subsection{Distribution of Similarity Scores}
\begin{figure}[h!] \centering \begin{subfigure}[b]{0.48\textwidth}
\vspace{-2mm}
\centering
\includegraphics[width=\textwidth]{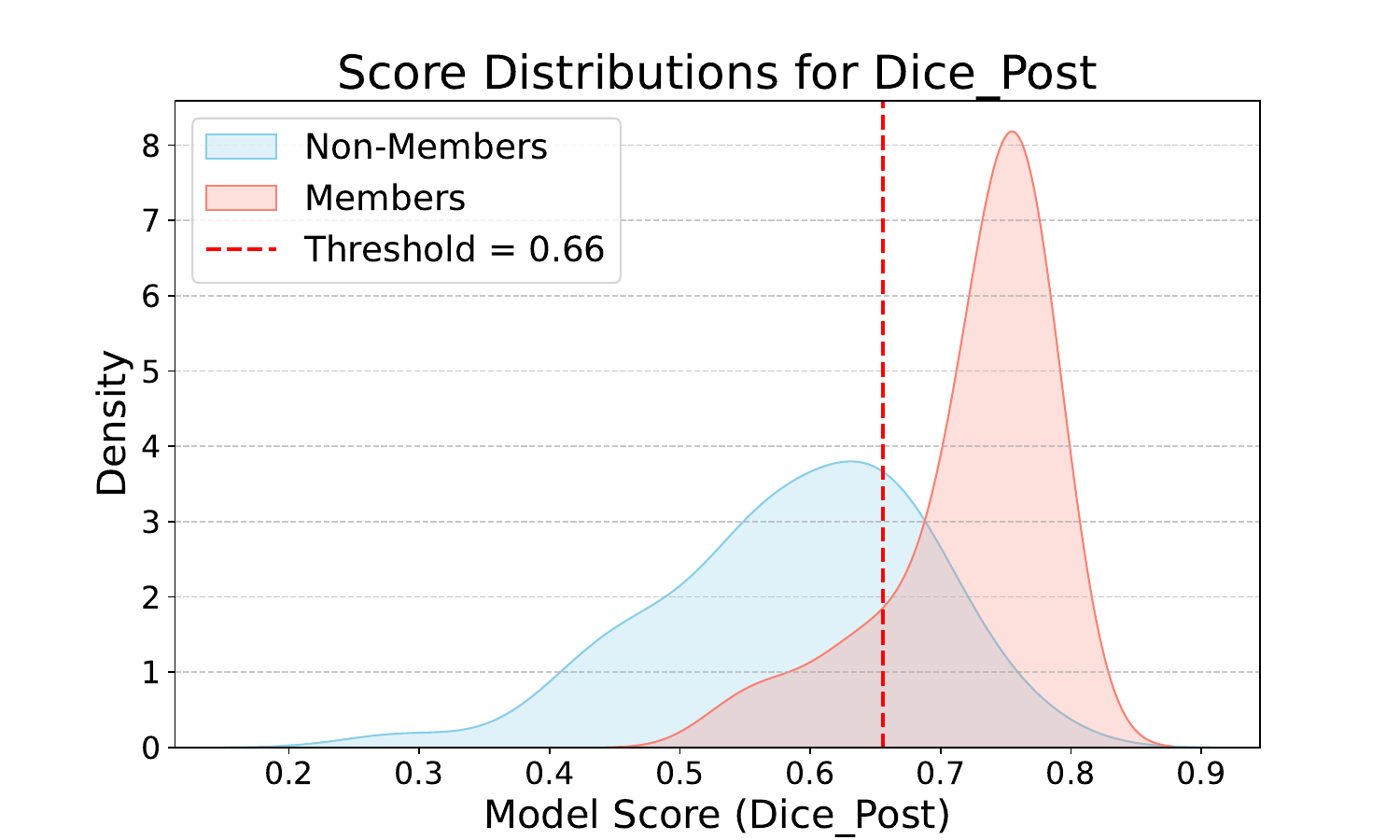}
\caption{Metal Layer as Member}
\label{fig:dist_metal}
\end{subfigure}
\hfill
\begin{subfigure}[b]{0.48\textwidth}
\centering
\includegraphics[width=\textwidth]{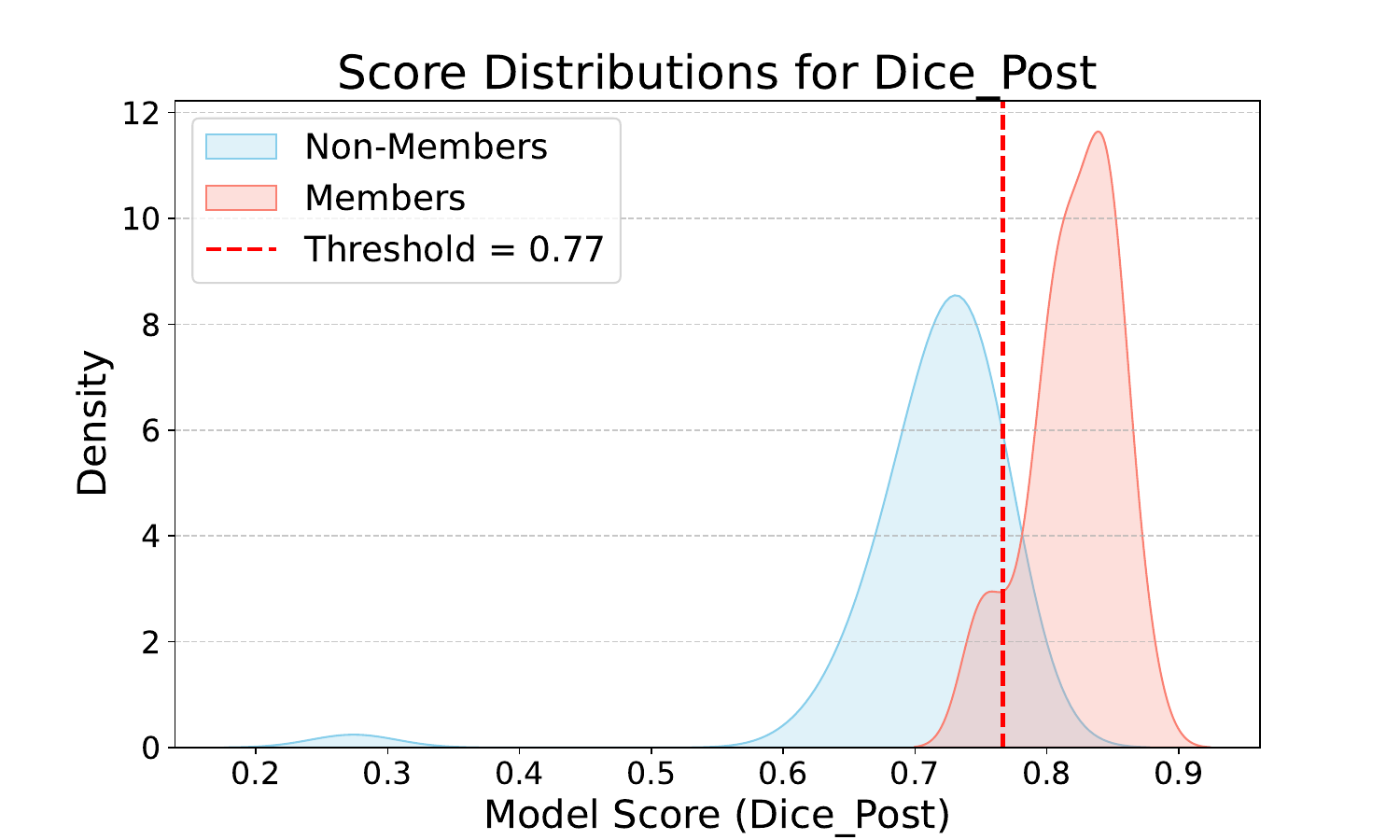} \caption{Diffusion Layer as Member}
\label{fig:dist_diffusion}
\end{subfigure}
\caption{Dice score distributions for members versus non-members at $\lambda_{\text{dummy}} = 0$.}
\label{fig:distributions}
\vspace{-3.5mm}
\end{figure}

To obtain quantitative metrics, we converted the raw reconstructions into binary masks using the post-processing pipeline outlined in Sect.~\ref{sec:methods} and computed the Dice similarity coefficient.
The results in Table~\ref{tab:dice_scores} validate that post-processed reconstruction quality serves as a reliable proxy for membership. As shown in Fig.~\ref{fig:distributions}, the mean Dice score was markedly higher when the guiding SCLL matched the specific category of the target's private data.

\subsection{Membership Inference Performance}
\begin{figure}[ht]
\vspace{-2mm}
\centering
\includegraphics[width=1\linewidth]{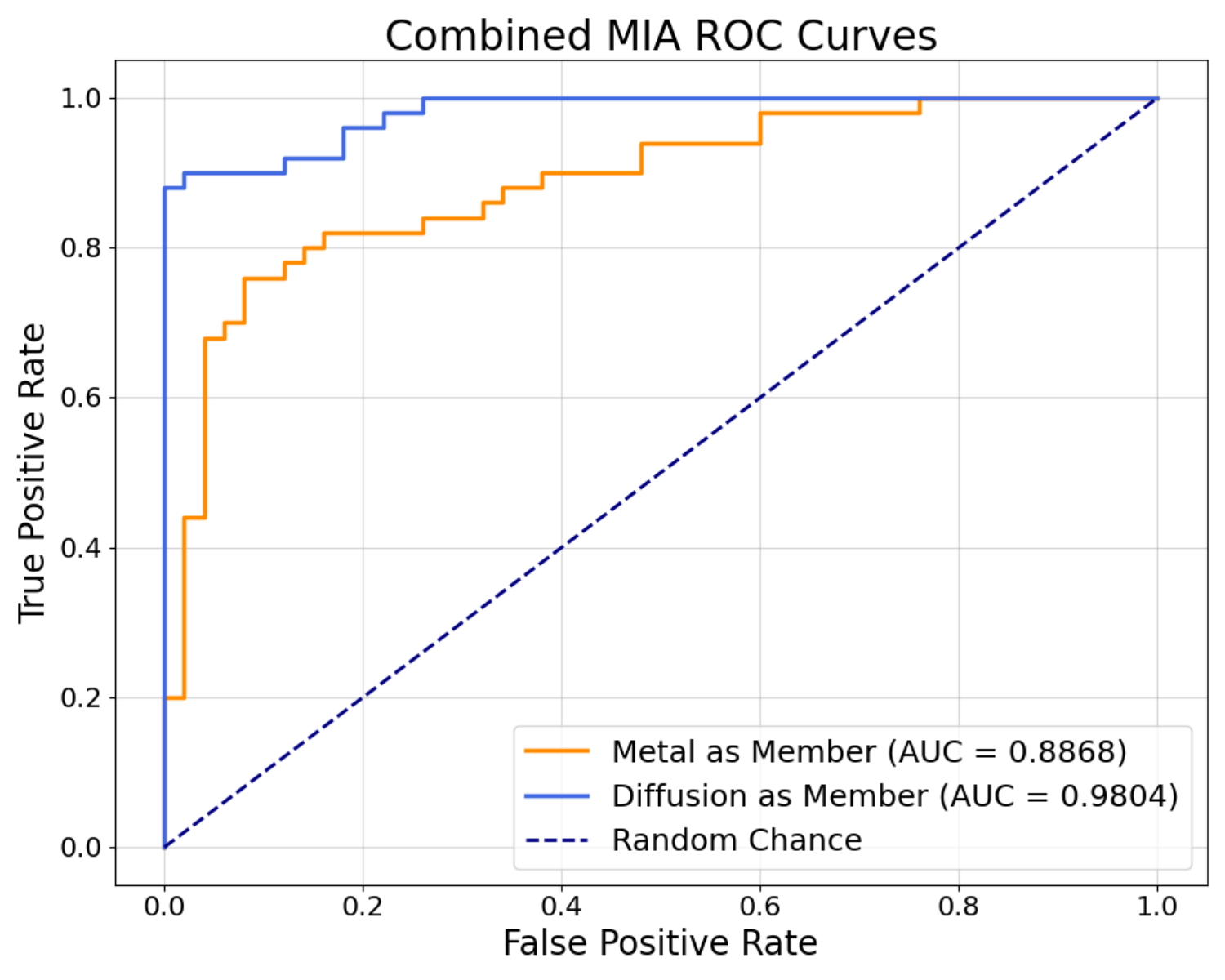} \caption{ROC curves for the MIA on metal vs. diffusion layers.}
\label{fig:roc_inter_layer}
\vspace{-3mm}
\end{figure}

We executed the membership inference attack utilizing the pooled similarity scores. The performance metrics are detailed in Table~\ref{tab:mia_performance}, with corresponding ROC curves displayed in Fig.~\ref{fig:roc_inter_layer}.
Although the attack succeeds in both scenarios, we observe a notable performance asymmetry: the diffusion layer yields a near-perfect AUC of 0.9804, surpassing the metal layer's AUC of 0.8868. This outcome is counter-intuitive, given that the raw reconstructions (Fig.~\ref{fig:reconstructions}) for the metal-as-member scenario visually display a more distinct quality gap. This discrepancy is likely attributed to a bottleneck in post-processing, as the intricate structures of the metal layer are highly susceptible to degradation during standardized blurring and thresholding. These operations can erode the fine lines of a high-fidelity metal reconstruction, artificially reducing its Dice score and limiting the threshold-based separation.

\begin{table}[t]
\centering
\caption{Metal vs. Diffusion Layers ($\lambda_{\text{dummy}} = 0$)}
\label{tab:mia_performance}
\begin{tabular}{@{}lcccc@{}}
\toprule Member Data & Accuracy & Precision & Recall & AUC \\
\midrule Metal Layer & 78.00\% & 0.75 & 0.84 & 0.8868 \\
Diffusion Layer & 94.00\% & 0.9783 & 0.90 & 0.9804 \\
\bottomrule
\end{tabular}
\vspace{-3mm}
\end{table}

\subsection{Ablation Study: Impact of $L_{\text{dummy}}$}
The negative Forward Pass loss term ($L_{\text{dummy}}$) was introduced as a regularizer to force the optimizer to prioritize gradient matching over fitting the fixed dummy label. Our results (Table~\ref{tab:ablation}) reveal that this mechanism is critical for structurally complex data.

For the complex metal layer, the standard GIA ($\lambda_{\text{dummy}} = 0$) results in a low AUC of 0.8868 due to the post-processing bottleneck described above. 
\begin{figure*}[ht!]
\centering
\includegraphics[width=0.9\linewidth, height=0.3\linewidth]{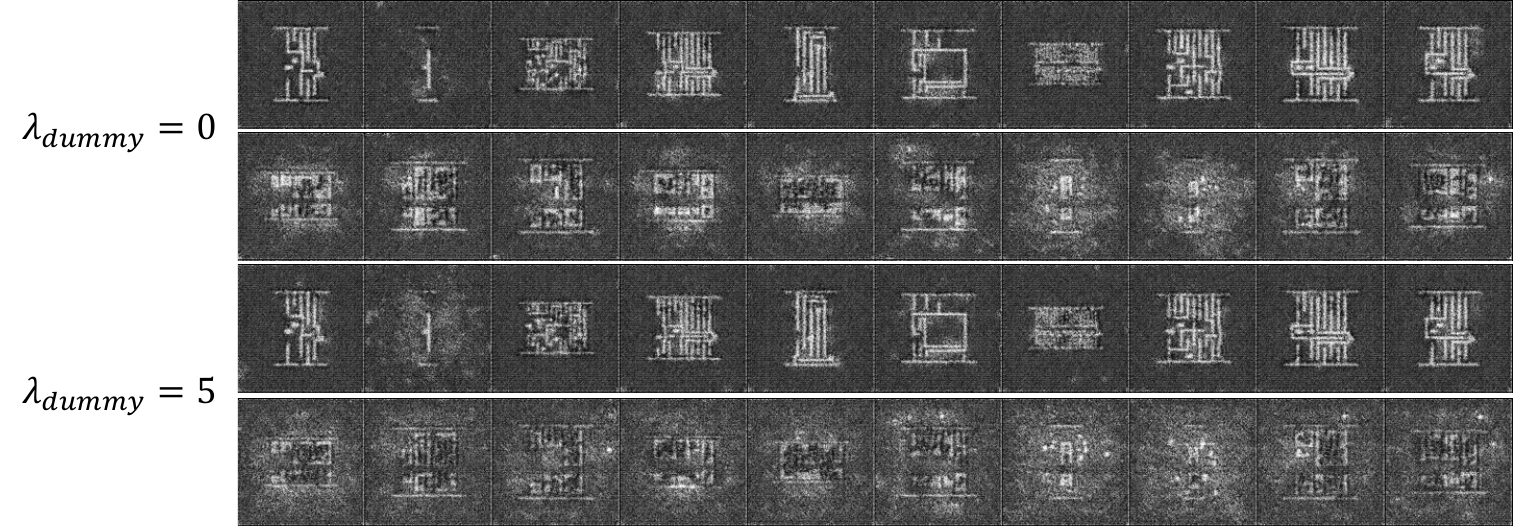} \caption{Visual influence of the $L_{\text{dummy}}$ term on metal targets. The quality disparity is visibly amplified at $\lambda_{\text{dummy}}=5$}
\label{fig:ablation_reconstructions}
\vspace{-2mm}
\end{figure*}
However, implementing $\lambda_{\text{dummy}}=5$ expands the measurable quality margin by disproportionately corrupting non-member reconstructions (see Fig. \ref{fig:ablation_reconstructions}). This enhancement raises the AUC to an optimal 0.9752. In contrast, the simpler diffusion layer inherently possesses a clear gradient signal, rendering this additional regularization superfluous. These results strongly demonstrate that the loss term functions primarily to reinforce the authentic gradient signal during complex evaluation scenarios.

\begin{table}[t]
\centering
\caption{Ablation Study on the FP Loss Term (AUC)} \label{tab:ablation}
\begin{tabular}{@{}ccc@{}} \toprule $\lambda_{\text{dummy}}$ & Metal Layer (Member) & Diffusion Layer (Member) \\ \midrule \textbf{0 (Term Removed)} & 0.8868 & \textbf{0.9804 (Optimal)} \\ 1 & 0.9376 & 0.9776 \\ 3 & 0.9720 & 0.9516 \\ 5 & \textbf{0.9752 (Optimal)} & 0.9228 \\ \bottomrule
\end{tabular}
\end{table}

\subsection{Intra-Layer Attack Performance}
We extended our evaluation to the challenging intra-layer task: distinguishing between 32nm and 90nm diffusion nodes. The results (Table~\ref{tab:intra_layer}) substantiate our bottleneck hypothesis. The intricate 32nm node, similar to the metal layer, necessitates $\lambda_{\text{dummy}}=5$ to reach a peak AUC of 0.9916. Conversely, the simpler 90nm node achieves optimal performance without this regularization. These findings confirm that the attack's ideal configuration is deterministic, governed by the interplay between the data's structural complexity and the constraints of the post-processing pipeline.

\begin{table}[t]
\centering
\caption{MIA Performance on Intra-Layer Datasets} \label{tab:intra_layer}
\resizebox{\columnwidth}{!}{%

\begin{tabular}{@{}lccccc@{}}
\toprule
Member Data & $\lambda_{\text{dummy}}$ & Accuracy & Precision & Recall & AUC \\
\midrule
32nm Diffusion & 0 & 67.00\% & 0.6308 & 0.82 & 0.6688 \\
\textbf{32nm Diffusion} & \textbf{5 (Optimal)} & \textbf{93.00\%} & \textbf{0.8909} & \textbf{0.98} & \textbf{0.9916} \\
\textbf{90nm Diffusion} & \textbf{0 (Optimal)} & \textbf{69.00\%} & \textbf{0.6610} & \textbf{0.78} & \textbf{0.8072} \\
90nm Diffusion & 5 & 64.00\% & 0.6207 & 0.72 & 0.7100 \\
\bottomrule
\end{tabular}
}
\vspace{-3mm}
\end{table}

\section{Discussion}
\label{sec:discussion}
Our research offers a more nuanced perspective on MIAs within Federated Learning. We establish that the attack's success is highly conditional, driven by the critical interaction between the intrinsic complexity of the private data and the evaluation methodology for attack.

A key contribution of this work is the identification of a post-processing bottleneck within the evaluation pipeline. We discovered that the conversion of raw reconstructions into binary masks for scoring unfairly penalizes complex, fine-grained structures compared to simpler ones. This bottleneck accounts for the initial underperformance of high-quality reconstructions (e.g., 32nm metal layers), as the evaluation metrics failed to accurately reflect their true fidelity.

This insight redefines the role of our loss term, $L_{\text{dummy}}$, as a strategic tool to effectively bridge this evaluation gap. By constraining the optimization to rely strictly on the gradient, the term induces a degradation in the raw quality of both member and non-member outputs. Crucially, this degradation is disproportionately severe for non-members, which lack a valid gradient signal to anchor the reconstruction. Consequently, this disproportionate corruption widens the measurable quality margin between the classes, facilitating a distinct separation of scores and significantly enhancing attack success.

\section{Future Works}
Our findings highlight several compelling directions for subsequent research, primarily centering on refining the attack methodology and addressing the challenges found in this study.

\begin{enumerate}
    \item \textbf{Refining Evaluation Metrics:} The identification of the post-processing pipeline as a performance bottleneck is a key insight. Future work should prioritize the development of advanced evaluation frameworks. Potential solutions include designing adaptive post-processing pipelines that automatically modulate parameters in response to the estimated complexity of the reconstruction.

    \item \textbf{Autonomous Hyperparameter Tuning:} We observed that the optimal attack configuration, particularly the $\lambda_{\text{dummy}}$ term, is data-dependent. A significant avenue for future work is the development of auto-tuning mechanisms for this parameter. This would augment the attack's universality, enabling effective deployment across diverse datasets without the need for manual calibration or prior knowledge.

    \item \textbf{Scalability and Real-World Applicability:} While our current experiments utilized a two-client framework, real-world FL deployments are extensive. To better approximate realistic threats, subsequent studies should scale the simulation to involve a larger cohort of clients and investigate the attack's performance against more complex backbones (e.g., Vision Transformers) beyond the standard CNN-based U-Net used in this study.

    \item \textbf{Assessing Countermeasures:} Given the demonstrated potency of the attack, a vital next step is the comprehensive evaluation of defenses. We have categorized applicable privacy techniques in Table~\ref{benefits_drabacks_table}, noting their specific utility against our GIA-based MIA in hardware assurance. Future research should rigorously benchmark these mechanisms, analyzing the privacy-utility trade-offs in perturbation methods like Differential Privacy and Synthetic Data and the computational feasibility of cryptographic protocols such as Homomorphic Encryption and Secure Multiparty Computation.
\end{enumerate}

\begin{table}[h!]
    \centering
    \footnotesize
    \setlength{\tabcolsep}{2pt}
    \begin{tabularx}{\columnwidth}{|>{\raggedright\arraybackslash}X|>{\raggedright\arraybackslash}X|>{\raggedright\arraybackslash}X|>{\raggedright\arraybackslash}X|}
        \hline
        \textbf{Techniques} & \textbf{Benefits} &\textbf{ Drawbacks} & \textbf{Examples} \\
        \hline
        Differential Privacy & Prevents high-fidelity reconstruction (GIA) by masking gradient details & Noise degrades segmentation precision for critical layers (e.g., Metal, Poly, Via) & DP-SGD on layout gradients \\
        \hline
        Synthetic Data & Replaces sensitive IP (layouts) with fake data, rendering reconstruction useless & Generative models may memorize sensitive training samples, leaking original IP & Training on GAN-generated circuit layouts \\
        \hline
        Secure Multiparty Computation & Aggregates updates securely; prevents access to individual gradients required for GIA & High communication overhead for large layout datasets and models & Secure Aggregation (SecAgg) of weights \\
        \hline
        Homomorphic Encryption & Gradients remain encrypted during aggregation, blocking adversary access & Prohibitive computational cost for deep CNNs used in vision-based assurance & Fully Homomorphic Encryption (FHE) training \\
        \hline
        Zero-Knowledge Proof & Verifies local training correctness without revealing the gradient update itself & High computational complexity to generate proofs for neural networks & ZK-SNARKs for update validity \\
        \hline
    \end{tabularx}
    \caption[Overview of defenses against reconstruction attacks]{Overview of privacy-preserving techniques against reconstruction MIAs in hardware assurance.}
    \label{benefits_drabacks_table}
    \vspace{-4mm}
\end{table}

\section{Conclusion}
In this paper, we expose a critical privacy vulnerability in Federated Learning, questioning its assumed privacy preservation. We prove that an adversary can successfully achieve higher success rate of MIA without auxiliary data, overcoming even a biased evaluation framework.

A primary contribution of DECIFR is the identification of a significant post-processing bottleneck. We discovered that this limitation, inherent to reconstruction-based attacks, obscures actual leakage by penalizing complex data structures. We resolve this issue by augmenting the attack with a negative Forward Pass loss term ($L_{\text{dummy}}$). This term serves as a conditional amplifier, establishing a clear quality margin by disproportionately degrading non-member reconstructions.

Ultimately, this discovery represents a profound threat to FL-based hardware assurance. By inferring specific hardware metadata (e.g., technology nodes or layers) instead of mere data points, adversaries can expedite physical threats such as reverse engineering, side-channel attacks, and fault injection.


\balance

\bibliographystyle{IEEEtran}
\bibliography{IEEEabrv,GOMACTech_LaTeX}

\end{document}